# MATAWS: A Multimodal Approach for Automatic WS Semantic Annotation

Cihan Aksoy[1,2], Vincent Labatut[1], Chantal Cherifi[1,3] and Jean-François Santucci[3],

[1] Galatasaray University, Computer Science Department, Ortaköy/İstanbul, Turkey
[2] TÜBİTAK, Gebze/Kocaeli, Turkey
[3] University of Corsica, Corte, France
caksoy@uekae.tubitak.gov.tr

**Abstract.** Many recent works aim at developing methods and tools for the processing of semantic Web services. In order to be properly tested, these tools must be applied to an appropriate benchmark, taking the form of a collection of semantic WS descriptions. However, all of the existing publicly available collections are limited by their size or their realism (use of randomly generated or resampled descriptions). Larger and realistic syntactic (WSDL) collections exist, but their semantic annotation requires a certain level of automation, due to the number of operations to be processed. In this article, we propose a fully automatic method to semantically annotate such large WS collections. Our approach is multimodal, in the sense it takes advantage of the latent semantics present not only in the parameter names, but also in the type names and structures. Concept-to-word association is performed by using Sigma, a mapping of WordNet to the SUMO ontology. After having described in details our annotation method, we apply it to the larger collection of real-world syntactic WS descriptions we could find, and assess its efficiency.

**Keywords:** Web Service, Semantic Web, Semantic Annotation, Ontology, WSDL, OWL-S.

## 1 Introduction

The semantic Web encompasses technologies which can make possible the generation of the kind of intelligent documents imagined ten years ago [1]. It proposes to associate semantic metadata taking the form of concepts with Web resources. The goal is to give a formal representation of the meaning of these resources, in order to allow their automatic processing. The process of defining such associations is known as semantic annotation (or annotation for short), and generally relies on libraries of concepts collectively described and structured under the form of ontologies. The result is Web documents with machine interpretable mark-up that provide the source material for software agents to operate. The annotation of Web resources is obviously fundamental to the building of the semantic Web.

According to Nagarajan and Uren *et al.*, in order to properly treat documents, annotating systems must follow a generic process [2] and meet seven different requirements [3]. The annotation process is composed of three primary steps that are the identification of the entity to be annotated, its possible disambiguation and its association to a concept. The requirements are as follow. The first one is to use standard formats (R1). Indeed, they provide a bridging mechanism that allows the access to heterogeneous resources and collaborating users and organizations to share annotations. The second one is to provide a single point of entry interface (R2), so that the environment in which users annotate documents is integrated with the one in which they create, read, share and edit them. The third one is to support multiple ontologies and to cope with changes made to ontologies (R3). This last point ensures consistency between ontologies and annotations with respect to ontology changes. The fourth and the fifth requirements are related to the document to be annotated. An annotating system must support heterogeneous input formats (R4), and be able to manage the annotation consistency when the document evolves (R5). The sixth requirement is about the annotation storage (R6), for which two models are proposed: the annotations can be stored separately from the original document or as an integral part of the document. Seventh, and finally, as manual semantic annotation leads to a knowledge acquisition bottleneck, the last requirement deals with the automation of the annotating process (R7). Automated annotation provides the scalability needed to annotate existing documents on the Web, and reduces the burden of annotating new documents.

Besides static Web content such as textual or multimedia documents, semantic annotation also concerns dynamic content, and more particularly Web Services (WS). WS are non-static in nature; they allow carrying out some task with effects on the Web or the real-world, such as the purchase of a product. The semantic Web should enable users and agents to discover, use, compose, and monitor them automatically. As Web resources, classic WS descriptions such as WSDL files can be semantically enhanced using the annotation principle we previously described, i.e. by the association of various ontological concepts. However, due to the particular structure of WS descriptions, these associations must comply with very specific constraints, which are different from those encountered for other kinds of Web resources such as Web pages. [2]. Indeed, the semantics associated with WS need to be formulated in a way that makes them useful to the application of WS. Sheth presents four types of semantics for the complete life cycle of a Web process: data, functional, non-functional and execution [4]. *Data* semantics is related to the formal definition of data input and output messages. *Functional* semantics is related to the formal definition of WS capabilities. *Non-functional* semantics is related to the formal definition of constraints like QoS. *Execution* semantics is related to the formal definition of execution flows at the level of a process or within a WS. Semantically annotating a WS implies describing the exact semantics of the WS data and functionality elements, which are crucial for the use of the WS, as well as its non-functional and execution elements.

Efforts for WS annotation include WS semantic languages as well as tools to annotate legacy WSDL files. The most prominent semantic languages are OWL-S [5],

WSMO [6], WSDL-S [7] and SAWSDL [8]. While OWL-S and WSMO define their own rich semantic models for WS, WSDL-S and SAWSDL work in a bottom-up fashion by preserving the information already present in WSDL. Those description languages are used in many research projects focusing on various semantic-related applications like automatic discovery and composition. In order to test these applications, one needs a benchmark, i.e. a large collection of annotated WS [9]. Such collections exist, but are limited in terms of size, realism, and representativity. These limitations are due to the fact the annotation process is generally performed manually, and is therefore costly. The use of an appropriate annotation tool can help decrease this cost, especially if it is automated. However, because of the specific structure of this kind of document, automatically annotating a WS description is much different, from the natural language processing perspective, than annotating other Web documents such as plain text. It consequently requires to perform a particular form of text mining, leading to dedicated tools such as ASSAM [10] or MWSAF [11]. But those tools also have their own limitations, the main one being they are only partially automated and require human intervention, which is a problem when annotating a large collection of WS descriptions.

In this paper we present the first version of MATAWS (Multimodal Automatic Tool for the Annotation of WS), a new semantic WS annotator, whose purpose is to solve some of these limitations. MATAWS was designed with the objective of batch annotating a large collection of syntactic descriptions and generating a benchmark usable to test semantic-related approaches. It focuses on data semantics (i.e. the annotation of input and output parameters) contained in WSDL files, and currently generates OWL-S files (other output formats will shortly be included). Our main contributions are: 1) a full automation of the annotation process and 2) the use of a multimodal approach. We consider not only the parameter names, but also the names present in the XSD types used in the WSDL descriptions: type names, and names of the fields defined in complex types.

The rest of this paper is organized as follows. Section 2 presents both existing ways of retrieving a collection of semantic WS descriptions: recover a publicly available collection and annotate a syntactic collection using one of the existing annotation tools. In section 3, we introduce MATAWS and describe our multimodal approach. In section 4 we apply MATAWS to the annotation of a publicly available collection of syntactic WS descriptions. Finally, in section 5 we discuss the limitations of our tool and explain how we plan to solve them.

## 2 Solutions to Access an Annotated Collection

When looking for a collection of semantic WS descriptions, one can consider two possibilities: either using a predefined collection, or creating his own one. In this section, we first review the main existing collections and their properties. The creation of a collection can be performed either by using a random model to generate artificial descriptions, or by semantically annotating a collection of real-world syntactical descriptions. The usual goal when looking for a semantic collection is to test WS-related tools on realistic data. To our opinion, the WS collections properties are not known well enough to allow the definition of a realistic generative model, which is why we favor the second solution. For this reason, in the second part of this section, we also review the main tools allowing to annotate WS descriptions.

### 2.1 Collections of Semantic Descriptions

The main publicly available collections of semantic WS are those provided by the ASSAM WSDL Annotator project, SemWebCentral and OPOSSum. Their major features are gathered in Table 1.

The ASSAM WSDL Annotator project (Automated Semantic Service Annotation with Machine learning) [12] includes two collections of WS descriptions named *Full Dataset* and *Dataset2*. *Full Dataset* is a collection of categorized WSDL files, which contains 816 WSDL files describing real-world WS. *Dataset2* is a collection of OWL-S files, obtained by annotating a subset of the WSDL files using the ASSAM Annotator (cf. section 2.2). 164 descriptions were fully labeled, assigning ontology references to the WS itself, its operations and their inputs and outputs.

**Table 1.** Collections of semantic WS descriptions: main features.

| Name | Dataset2 | OWLS-TC3 | SAWSDL-TC | SWS-TC |
|---|---|---|---|---|
| Source | ASSAM project | SemWeb Central | SemWeb Central | SemWeb Central |
| Type | Real-world descriptions | Real-world descriptions, partially resampled | Real-world descriptions, partially resampled | N/A |
| Language | OWL-S | OWL-S | SAWSDL | OWL-S |
| Annotated Type | Data, Functional | Data | Data | Data |
| Size | 164 | 1007 | 894 | 241 |
| Particular features | Processed using Assam annotator | Single interface, one operation per service | Single interface, one operation per service | N/A |

SemWebCentral [13] is a community whose purpose is to gather efforts from people working in the semantic Web area. Three semantic collections are available: *OWLS-TC* (OWL-S Test Collection), *SAWSDL-TC* (SAWSDL Test Collection) and *SWS-TC* (Semantic WS Test Collection). OWLS-TC3 is the third version of this test collection. It provides 1007 semantic descriptions written in OWL-S from seven different domains. Part of the descriptions were retrieved from public IBM UDDI registries, and semi-automatically transformed from WSDL to OWL-S. SAWSDL-TC originates in the OWLS-TC collection. It was subsequently resampled to increase its size, and converted to SAWSDL. The collection provides 894 semantic WS descriptions. The descriptions are distributed over the same seven thematic domains than OWLS-TC. SWS-TC is a collection of 241 OWL-S descriptions. There is not much information about this collection.

OPOSSum (Online POrtal for Semantic Services) [14] is a joint community initiative for developing a large collection of real-world WS with semantic descriptions. Its aim is to create a suitable test bed for semantically enabled WS technologies. OPOSSum gathered the three semantic collections of SemWebCentral, plus the *Jena Geography Dataset* collection, explicitly collected within OPOSSum. The collection contains 201 real-world WS descriptions retrieved from public. All the described WS belong to the domains of geography and geocoding. Unfortunately, for now, no semantic descriptions are available for the services of the Jena Geography Dataset, which is why this collection is absent from Table 1.

These collections have been widely used in semantic WS-related works [15, 16]. As shown in Table 1, they all focus on the annotation of the data elements, which corresponds to our objective. However, one can notice some limitations. SWS-TC description is insufficient, it is not even clear if the WS descriptions are real-world. Dataset2 contains only real-world WS descriptions but it is very small, which can raise questions about its representativity. On the contrary, OWLS-TC3 and SAWSDL-TC contain a substantial number of descriptions. Nevertheless, these have been partially resampled in an undocumented way, which raises important questions concerning their realism.

## 2.2 Annotation Tools

From our point of view, WS annotation is considered as a one-time task, aiming at annotating legacy WS, which are described only syntactically. Newly created or modified WS should be (re)annotated manually by their authors, which is much preferable in terms of quality than any automatic processing. For this reason, and due to the specific nature of WS annotation, we are not concerned by all the 7 requirements stated by Uren *et al.* [3] for general annotation tools. It is of course necessary to use standard formats for input and output (R1). A polyvalent environment is not necessary, since we do not want to modify existing descriptions or create any new ones (R2). The support of multiple or changing ontologies is relevant (R3), but it is not the most important point, so we chose to ignore it in this first work. The input format is constrained to WSDL (R4), since it is the *de facto* standard for syntactical WS description. As stated before, we do not plan to maintain annotations if WS are modified (R5). The model of annotation storage (R6) is constrained by the output format: separate form for OWL-S and integrated for WSDL-S and SAWSDL. Finally, the level of automation is of great interest to us, given our context (R7).

Only a few publicly available tools exist to semantically annotate WS descriptions. Table 2 presents the main ones and summarizes their properties. They all take a set of WSDL files as input (R1 and R4), but differ on several properties such as their level of automation (R7) and the language used to output the semantic descriptions (R1). The tools are described in details in the rest of this subsection.

*Radiant* is an open source tool created at the Georgia University [17]. It takes the form of an Eclipse plug-in and can output both SAWSDL and WSDL-S files. It provides a GUI which presents the elements constituting the WS description and allows to select the concepts one wants to associate to parameters or operations, by browsing in the selected ontologies. This interface makes the annotation process easier, but the annotation is nevertheless fully manual.

*ASSAM* is an open source Java program developed at the University College Dublin [12], able to output OWL-S files. It provides assistance during the annotation process. First, the user starts manually annotating parameters and/or operations using an existing ontology. Meanwhile, ASSAM identifies the most appropriate concepts using machine learning methods. After enough information has been provided, the software is able to propose a few selected and supposedly relevant concepts when the user annotates a new WS.

*MWSAF* is another open source Java tool created at the Georgia University [11]. It outputs WSDL-S files, and like ASSAM it has a machine learning capability allowing it to assist the user during the annotation process. It is able to annotate not only parameters and operations, but also non-functional elements.

*WSMO* Studio is an Eclipse plug-in initially designed to edit semantic WS based on the WSMO model. An extension allows annotating WS parameters and operations, and outputting the result under the form of SAWSDL files [18]. However, the tool does not provide any assistance to the user and the process is fully manual.

**Table 2.** WS Semantic annotation tools and their properties.

| Name | Output Format | Annotated Type | Automation | Last Update |
|---|---|---|---|---|
| Radiant | SAWSDL, WSDL-S | Data, Functional | Fully manual | May 2007 |
| ASSAM | OWL-S | Data, Functional | Assisted | May 2005 |
| MWSAF | WSDL-S | Data, Functional, Non-Functional | Assisted | July 2004 |
| WSMO Studio | SAWSDL | Data, Functional | Fully manual | Sept. 2007 |

Besides these annotation tools, several softwares allow to convert WSDL files to OWL-S files, but without performing any semantic annotation: they only apply a syntactic transformation and present the information contained in the original WSDL file under a form compatible with the OWL-S recommendation. *WSDL2OWLS* is an open source Java application created at the Carnegie Mellon University [19]. *OWL-S Editor* is a plug-in for Protégé (itself an ontology development environment) created at SRI [20]. Another software performing the same task is also called *OWL-S Editor*, but was developed at Malta University [21].

From this review, we can conclude the existing annotation tools present various limitations relatively to our goals. First, from a practical perspective, some of these tools are old and not supported anymore, which can cause installation and/or use problems. For instance, Radiant and ASAM are not compatible anymore with the current versions of some of the Eclipse plug-ins, libraries or API they rely on; meanwhile MWSAF installs and runs fine, but generates files without any of the annotations defined by the user. More importantly, these tools require important human intervention: Radiant and WSMO Studio are fully manual, whereas ASSAM and MWSAF only assist the user, after a compulsory learning phase. This justifies the development of our own tool, which we present in the next section.

## 3  Proposed Annotation Method

The absence of an existing solution fulfilling our needs compelled us to develop our own tool to semantically annotate WS descriptions. The main differences with the other annotation tools are the exploitation of several sources of information and the automation of the annotation process. In this section, we first describe the general architecture of our tool, which is made up of several independent components. We then focus separately on the components of interest, explaining their design and functioning.

### 3.1  General Architecture

MATAWS takes a collection of WSDL files as input and generates a collection of OWL-S files as output. Fig. 1 gives an insight of its modular structure, which includes five different components. Among these components, two are using external APIs (Associator and Output Component), whereas the three remaining ones were developed by us in Java. The Input and Output components are not of great interest with regards to the topic of this article, which is why we describe them shortly here. The other components are described in details in the following subsections.

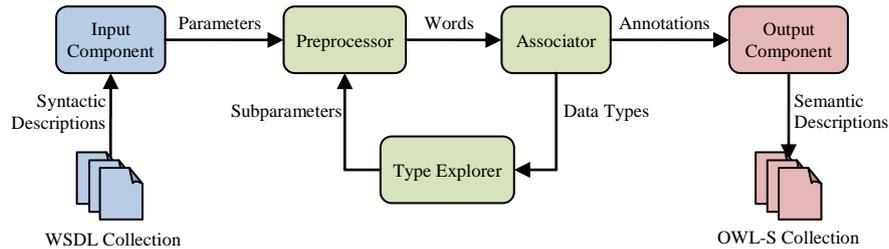

**Fig. 1.** Architecture of MATAWS.

The *Input Component* is in charge for extracting the set of all operation parameters defined in the considered collection of WSDL files. We designed a parser able (among other things) to retrieve the parameter names, type names and type structures (in the case of complex types) [22]. The *Output Component* is used after the annotation process to generate a collection of OWL-S files corresponding to annotated versions of the input WSDL files. For this purpose, we selected the Java *OWL-S API*, which provides a programmatic read/write access to OWL-S service descriptions [5]. Note we plan to add support for WSDL-S and SAWSDL by using other appropriate APIs.

The three remaining components correspond to the core of the annotation process. After the input component has parsed the WSDL files, it fetches parameters information to the *Preprocessor*. This one originally focuses on the parameter names, decomposing, normalizing and cleaning them so that they can be treated by the *Associator*. This component is based on the inference engine Sigma [23], whose role is to associate an ontological concept to a word. If Sigma is successful and manages to return a concept, this one is associated to the considered parameter. After all the parameters of a WS have been annotated, the *Output Component* is used to generate an OWL-S file with both the information contained in the original WSDL file and the selected concepts. However, for various reason explained later, it is not always possible for Sigma to find a suitable concept for every parameter. In this case, the *Type Explorer* accesses some properties related to the parameter data type, to obtain what we call subparameters. These are then fetched to the Preprocessor and the core processing starts again. In case of repeated annotation failure, this process can be repeated recursively until success or lack of subparameters.

### 3.2 Preprocessor

In order to work properly and propose a suitable concept, the Associator needs to process clear and normalized words. However, the names defined in real-world WS certainly do not meet this criterion. First, the meaning of an operation, parameter or type can hardly be described using a single word. For this reason, most names are made up of several concatenated words, separated either by alternating upper and lower cases or by using special characters such as dots, underscores, hyphens, etc. Second, sometimes the result is too long and abbreviations are used instead of the complete words. Finally, an analysis of any collection quickly shows different additional characters such as digits or seemingly useless separators can also appear.

Of course, there is no way to define an exhaustive list of the various forms a name can take in a WS description, but WS programmers actually follow only a few conventions, which allows performing very efficient preprocessing by applying a set of simple transformations to break a name into usable words. We distinguish three steps during name preprocessing: decomposition, normalization and filtering.

**Table 3.** Preprocessing examples.

| Transformation | Original Name | Extracted Words |
|---|---|---|
| Decomposition | `WhiteMovesNext` | White, Moves, Next |
| Decomposition | `Number3Format` | Number, Format |
| Decomposition | `AUsername` | Username |
| Decomposition | `User_name` | User, name |
| Normalization | `no` | number |
| Normalization | `Password` | password |
| Filtering | `Parameter` | - |
| Filtering | `Body` | - |

The decomposition consists in taking advantage of the different types of concatenations we identified to break a name into several parts. It also involves some cleaning, in the sense all characters which are not letters are removed and diacritical marks are deleted. Table 3 shows some examples involving case alternation, and digit and underscore used as separators.

The normalization role is first to provide the Associator a clean version of the word, typographically speaking, by setting each word to lower case. Moreover, the normalization handles abbreviations, by replacing them with the corresponding full-length words. Table 3 gives an example of the name `no` being replaced by the word *number*. However, this last task is very context-dependent, because some strings are both full words and common abbreviations. For instance, `no` could simply mean the opposite of "yes", used to negate the following concatenated word, e.g. `no_limit`. For this reason, human intervention can be necessary to set up this preprocessing, and adapt it to the considered collection. We chose to allow the user to define a list of common abbreviations.

Finally, we added a filtering step to deal with stop-words, i.e. words with no particular semantic information relatively to their context. For instance, the string `parameter` commonly appears in parameter names, without bringing any significant information, since the syntax of the WSDL file already allows to know if a certain name points out at a parameter. For this reason, it can be considered as noise and ignored. Even more than before, the nature of the stop-words is closely linked to the domain of application, and requires human intervention to adapt the list of stop-words we defined.

Let us consider as an example the preprocessing of the name `ASessionId_02`. First it will be broken down to the words *A*, *Session* and *Id* while the numeric end of the name (`02`) will be ignored. The normalization step will transform them in *a*, *session* (lowercase) and *identity* (replacing an abbreviation). Finally, the filter will remove the article *a*, because it is a stop-word. Eventually, for this name `ASessionId_02`, the Preprocessor will output the two words *session* and *identity*.

### 3.3   Associator

As mentioned before, we use an existing tool called Sigma to associate a concept to a word. It is written in Java and allows to create, test, modify and infer ontologies [23]. It is provided with the Suggested Upper Merged Ontology (SUMO), which (unlike its name suggests) contains also mid-level and domain ontologies [24]. SUMO is free, covers a wide range of fields, and it has been mapped to the whole WordNet lexicon [25]. It was initially defined using the SUO-KIF language [26], and it is currently being converted in OWL [27].

**Table 4.** Concept association examples.

| Word | SUMO Concept associated by Sigma |
|---|---|
| buffalo | `HoofedMammal` |
| school | `EducationalProcess` |
| talk | `Communication` |

Although its main purpose is to work on ontologies, Sigma also offers a programmatic access to this mapping under the form of a method taking an English word as input and outputting a SUMO concept. Table 4 gives a few examples of such associations. The names we processed are most of the time not plain English words, which justifies our preprocessing.

### 3.4 Type Explorer

Although our focus is primarily on parameter names, we described the two previous components in general terms, because they can be applied to any kind of names. Indeed, different difficulties can arise, making it impossible to associate a concept to a parameter name. First, the Preprocessor might fail to break the name down to relevant words, hence fetching the Associator strings it cannot map to appropriate concepts. Second, the Preprocessor might filter all the words resulting from the name decomposition, meaning it will not be able to provide the Associator any word to process. This can be the case, for instance, when a name is composed of a single stop-word or several concatenated ones (e.g.: `SomeParameter_08`). Third, even if at least one correct English word can be fetched to the Associator, it is possible this one simply does not find any associated concept.

All three cases, or any combination of these three cases, result in the fact no concept could be associated to the considered parameter. To overcome this problem, we propose a multimodal approach taking advantage of latent semantics contained in the data type information available through WSDL files. First, in real-world WS, a large proportion of types have a user-defined name, whose meaning can be considered as complementary to the parameter name. Additionally, many of these custom types are complex in the XSD sense, i.e. they can be compared to the structured data types used in programming languages. A parameter whose type is complex is made up of several subparameters, which can recursively be composed themselves of other subparameters, if they have a complex type too. Therefore, by taking advantage of the data types, one can access the semantic information implicitly contained in the type names and subparameter names and types.

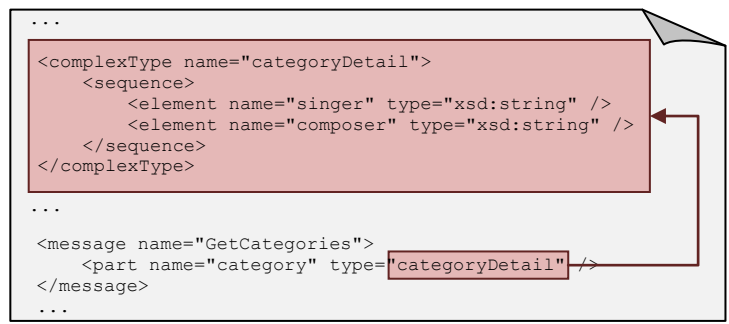

**Fig. 2.** Excerpt from a real-world WSDL file: parameter with a complex XSD type.

Fig. 2 gives an example of a complex type extracted from a real-world WSDL file. A parameter named `category` has a complex type called `categoryDetail`, defined as a sequence of two strings: a `singer` and a `composer`. If we suppose the word

*category* is a stop-word, the Associator will not be able to provide any concept for this parameter. However, considering the words *singer* and *composer* gives access to additional information usable by the Associator.

The principle of our Type Explorer component is as follows. It is activated when the processing of the parameter name could not be used to successfully identify any concept. We start with the type name: if it is custom, we process it exactly like the parameter name, going through the preprocessing and association steps. In case of failure to associate any concept, we go further and consider the type structure. If it is complex, we access the first level of subparameters. For now, we only consider XSD sequences, because these are the most widespread, however the same approach can be extended to the other kinds of XSD types. We first focus on the subparameter names, and if the association is inconclusive, on their type names. In case of failure, the process recursively goes on by analyzing the structure of the subparameter types to access the second level of subparameters. The recursion stops when there is no more level to process (permanent failure) or as soon as concept can be associated (success).

## 4      Application to Real-World Descriptions

To assess its performance, we applied MATAWS to a collection of syntactic WS descriptions. We wanted to use a large collection of real-world descriptions, in order to avoid specific cases and to get consistent results. Given these criteria, the best collection we could found is the *Full Dataset* collection from the ASSAM project [12], previously mentioned in our review of WS descriptions collections (section 2.1). It contains 7877 operation distributed over 816 real-world WS descriptions. In this section, we present the results we obtained on this collection. First we adopt a quantitative point of view and distinguish parameters only in terms of annotated or non-annotated. Second, we analyze the results qualitatively and discuss the relevance of the concept associated to the parameters.

### 4.1     Quantitative Aspect

We first focus on the proportion of parameters from the Full Dataset collection which could be automatically annotated by MATAWS. In this section, we consider a parameter to be successfully annotated if our tool was able to associate it to at least one concept. Table 5 displays several values, corresponding to the progressive use of the different components described in section 3. Each row represents the performance obtained when using simultaneously the specified functionality and those mentioned in the previous rows.

The first line corresponds to the direct application of the Associator, with no significant preprocessing. The only transformation consists in setting parameter names to lowercase, which is compulsory to apply Sigma. Under these conditions, MATAWS can propose a concept for 39.63% of the parameters. This means close to 40% of the parameters names are single words, which can be retrieved directly in WordNet. The rest needs more preprocessing to be successfully annotated.

The second row corresponds to the introduction of the decomposition step. The small improvement in the success rate (around +2%) allows us to think compound names do not contain directly recognizable words. By adding the normalization step, the improvement is extremely large (almost +48%). Further analysis shows this is only marginally caused by the replacement of abbreviations by full words. Among the remaining 10%, one can found specific parameter forms we plan to introduce in our preprocessing, and word variations such as plural forms, also easily integrable in our approach.

**Table 5.** Success rates obtained by using the different functionalities of MATAWS.

| Added Modification | Proportion of Annotated Parameters |
|---|---|
| No preprocessing | 39,63% |
| Decomposition | 41,94% |
| Normalization | 90,01% |
| Filtering | 69,06% |
| Type Explorer | 72,04% |

A strong decrease (–21%) can be observed when introducing the filtering step. This means that, among the associated words, many correspond to stop-words, or concatenations of stop-words. In this case, the Annotator might be able to retrieve a concept, but this one is useless in this context (e.g. parameter). The introduction of the Type Explorer allows improving slightly our success rate (+3%), but its effect is not as strong as we expected. This can be justified by the fact most parameters with a custom type where annotated using only their names. Moreover, the type structure is difficult to exploit in this collection, because some types defined as complex surprisingly do not actually have any content (i.e. no subparameters at all).

### 4.2 Qualitative Aspect

The quantitative analysis reflects the fact a large proportion of parameters could be associated to a concept. The question is now to know if these associations, which were automatically retrieved, are relevant relatively to the context. For this matter, we isolated all the words detected in the whole set of parameters, thanks to our Preprocessor and Type Explorer. Table 6 shows the first most frequent words with their associated concept.

Overall, most of the annotated words are associated to relevant concepts, leading to an approximate success rate of 83%. Words like *computer*, *month*, *numeric*, *password*, *customer* are perfectly recognized, but this is not the case of several widespread words such as *name*, *user*, *address* or *value*.

Irrelevant concepts are due to the fact some words have several meanings and can therefore be associated to several concepts. Such ambiguity can be raised directly when the considered word has most probably a unique meaning in the context of WS. For instance, when the word *user* is submitted to Sigma, it outputs three concepts, including the one expected in this case, i.e. "someone employing something".

However, the top result corresponds to "someone who does drugs", which explains the associated concept (`DiseaseOrSyndrome`). Similarly, the appropriate concept for *name* is among the concepts returned by Sigma, but the top result correspond to its meaning in the expression "in the name of the law", hence the concept (`HoldsRight`). The quality of the annotation could be improved for such common words by simply selecting *a priori* the appropriate concepts, like we defined lists of stop-words and abbreviations.

**Table 6.** List of the most frequent words, with their associated concept. Bold rows represent semantically irrelevant concepts.

| Word | Occurrences | Associated Concept |
| --- | --- | --- |
| identity | 1255 | `TraitAttribute` |
| key | 548 | `Key` |
| **name** | **470** | **`HoldsRight`** |
| **user** | **424** | **`DiseaseOrSyndrome`** |
| **code** | **295** | **`Procedure`** |
| **number** | **294** | **`Object`** |
| **address** | **258** | **`SubjectiveAssessmentAttribute`** |
| **date** | **203** | **`DateFruit`** |
| city | 168 | `City` |
| amount | 135 | `ConstantQuantity` |
| administrator | 128 | `Position` |
| message | 115 | `Text` |
| **value** | **106** | **`ColorAttribute`** |
| password | 98 | `LinguisticExpression` |
| pass | 70 | `ContestAttribute` |
| customer | 52 | `Customer` |
| company | 51 | `Corporation` |
| phone | 41 | `Device` |
| electronic | 35 | `ElectricDevice` |
| computer | 33 | `Computer` |
| mailing | 33 | `Transfer` |
| month | 32 | `Month` |
| numeric | 32 | `Number` |

The selection of an accurate concept can also be context dependent, which makes it impossible or difficult to perform it *a priori*. For instance, the word *value* corresponds to many concepts equally likely to appear in a WS description: quantity, monetary value, time duration, etc. Regarding this problem, the quality of the automatic annotation can be improved by deriving concepts from several words, when they are available. For instance, if the parameter name is `value01` and its type is `myCurrencyType`, then we have enough information to infer the most relevant concept. This can be done, for example, by taking advantage of the WordNet textual definitions.

## 5    Conclusion

In this article, we presented our tool MATAWS, which implements a new method to semantically annotate WS descriptions. It focuses on WS parameters, i.e. on the Data semantics [4], and implements most of the requirements defined by Uren *et al*. [3] and relevant to our context: it processes WSDL files and produces OWL-S files (R1 & R4), and is fully automated (R7). This automation level is enforced through the use of both an ontological mapping of the WordNet lexicon, and a multimodal approach consisting in using not only parameter names, but also data type names and structures to identify appropriate ontological concepts. When compared to existing annotation tools such as ASSAM [12] and MWSAF [11], it is important to notice that MATAWS is much less flexible, because it does not include any machine learning abilities. This is due to the fact our goal is different: we want to batch annotate a large collection of WS descriptions without any human intervention, whereas the cited works aim at helping human users to annotate individual WS descriptions. Moreover, we tested MATAWS on a large collection of syntactic real-world WS descriptions, and despite its simplicity, it obtained very promising results, with 72% of the parameters annotated.

The version presented in this article constitutes a first step in the development of our tool. Although some parameters could not be associated with relevant concepts, it is clear that we reduced the manual labor required for the annotation of WS. However, for now this reduction is not important enough to spare human intervention, which is needed at least to control the result of the annotation process. To get around this limitation, we plan to improve our tool on several points. First, in order to lower the proportion of parameters we failed to annotate, we can use other sources of latent semantics present in the WSDL descriptions: natural language descriptions and names of messages and operations. Second, the association step can be improved in two ways. We can complete the Associator by including more tools able to map a lexicon to an ontology, such as DBPedia [28]. This would complete and enhance the results already obtained through Sigma. Also, by taking advantage of our multimodal approach, we can retrieve all the words related to a given parameter through its data type, in order to compare them with concept definitions expressed in natural language (as found in a dictionary).

**Acknowledgments.** The authors would like to thank Koray Mançuhan, who participated in the development of MATAWS.